\documentclass[a4paper,11pt]{article}
\usepackage{pos}
\usepackage{epsfig}
\usepackage{bm}

\usepackage{amssymb}
\usepackage{fontenc}
\usepackage{times}
\usepackage{graphicx}

\title{Lattice QED in external electromagnetic fields}

\author*[a]{D.~K.~Sinclair}
\author[b]{J.~B.~Kogut}

\affiliation[a]{HEP Division, Argonne National Laboratory, 
9700 South Cass Avenue, Lemont,Illinois 60439, USA}

\affiliation[b]{Department of Energy, Division of High Energy Physics,
Washington, DC 20585, USA \\
and\\
Department of Physics -- TQHN, University of Maryland, 82 Regents Drive,
College Park, MD 20742, USA}

\emailAdd{dks@anl.gov}
\emailAdd{jbkogut@umd.edu}

\abstract{
We study QED in external electromagnetic fields using methods developed for
simulating lattice QCD. Our first project is to simulate QED in a constant
(in space and time) external magnetic field on a euclidean space-time lattice
using the Rational Hybrid Monte Carlo (RHMC) method. Observables we measure 
include the condensate $\langle\bar{\psi}\psi\rangle$ and the effective
electron action after integrating out the fermion fields. We look for evidence 
that the combined effect of the magnetic field and the electron-positron
attraction from QED produces a non-zero condensate in the limit of zero 
electron mass, a non-perturbative effect analogous to spontaneous chiral
symmetry breaking. Very preliminary evidence is that such a condensate exists,
at least for strong external magnetic fields and unphysically large electric
charge. In addition, we are storing field configurations to measure the 
expected distortions and screenings of the coulomb field of a charged particle
due to the vacuum polarization asymmetries produced by the magnetic field.
We hope also to measure the dynamical contribution to the electron mass
produced by the same mechanism that produces a finite condensate in the 
zero input mass limit.
}

\FullConference{%
 The 38th International Symposium on Lattice Field Theory, LATTICE2021
  26th-30th July, 2021
  Zoom/Gather@Massachusetts Institute of Technology
}

\begin{document}
\maketitle

\section{Introduction}

Theoretical studies of electrons in external electromagnetic fields date from
the early days of relativistic quantum mechanics \cite{Sauter,Heisenberg}
and exhibit some of the first features of quantum electrodynamics
\cite{Schwinger:1951nm} such as the Sauter-Schwinger effect -- the production
of electron-positron pairs in an external electric field. [See the review
article by Dunne \cite{Dunne:2012vv} for a good summary of the cases for which
exact, closed form (simple integral or series) solutions are known]. However,
such effects only become significant when the electric field $E$ is of order
$E_{cr}=m^2/e$ or larger where $m$ is the electron mass and $e$ is the electron
charge. Similarly external magnetic fields only produce significant quantum 
effects when the magnetic field $B$ is of order $B_{cr}=m^2/e$ or larger.
Interest in these effects has recently been revived by new and planned 
facilities with extremely intense laser beams, whose interactions with 
charged particle beams can produce such strong fields
\cite{Fedeli:2020fwt,Meuren:2021qyv}. Additionally it has been realized that
compact astronomical X-ray and $\gamma$-ray sources are probably neutron stars
with such strong fields. For a review, see for example \cite{Harding:2006qn}.
Finally it has been noted that in the next generation of electron/positron
colliders, beam-beam interactions can produce electromagnetic fields orders of
magnitude above $E_{cr}$ or $B_{cr}$, where all conventional QED calculations
break down \cite{Yakimenko:2018kih}.
     
We are interested in applying methods developed for simulating lattice QCD to
lattice QED in strong external electromagnetic fields. For QED in external
electric fields, the transition to euclidean space needed to apply lattice
simulation methods results in a complex action which prevents the use of
standard simulation methods. We are therefore starting with QED in external
magnetic fields where the transition to euclidean space produces a real 
action bounded below where standard simulation methods apply. 

We simulate lattice QED in a constant external magnetic field $B$ using the 
RHMC method of Clark and Kennedy \cite{Clark:2006fx}, with a non-compact gauge
action and staggered fermions, using rational approximations to tune to a
single electron. The coupling of the gauge fields to the electron is compact.
Not only does this ensure that the action is gauge invariant, but it also
allows an effectively constant magnetic field over the whole lattice.
Note that the internal gauge(photon) field is in both the gauge action and the
fermion action, while the external field is only in the fermion action.

Classically, charged particles in a constant magnetic field are restricted to
helical orbits around magnetic field lines. Quantum mechanics restricts the
motion of the charged particles in the plane transverse to the magnetic field
to a discrete set of evenly spaced levels -- the Landau levels -- while leaving
the motion in the direction of the magnetic field continuous. For a discussion
of Landau levels in relativistic quantum mechanics see for example
\cite{Akhiezer}. For strong magnetic fields, only low lying Landau levels are
populated resulting in the effective dimensional reduction from
$3+1$~dimensions to $1+1$~dimensions.

In quantum mechanics the chiral condensate $\langle\bar{\psi}\psi\rangle$ is
enhanced in the presence of the magnetic field, but still vanishes as 
$m \rightarrow 0$. Various approximate calculations suggest that the addition
of QED causes $\langle\bar{\psi}\psi\rangle$ to develop a non-zero contribution
proportional to $(eB)^{3/2}$ which does not vanish in this limit, and 
furthermore that the electron gains a non-zero dynamic mass proportional to
$\sqrt{eB}$, \cite{Gusynin:1994xp,Gusynin:1995nb,Gusynin:1999pq,Leung:1996poh,
Alexandre:2000nz,Alexandre:2001vu,Wang:2007sn}. This non-perturbative effect
is referred to as `magnetic catalysis'. For good review article with a more
complete set of references see Miransky and Shovkovy \cite{Miransky:2015ava}.

In QED, one effect of the magnetic field is to introduce asymmetries in the
vacuum polarization. These distort the coulomb field of any charged particle
in such an external magnetic field and also induce at least partial screening
\cite{Shabad:2007xu,Shabad:2007zu,Sadooghi:2007ys,Machet:2010yg}.
On the lattice, such effects can be measured using Wilson loops.

We simulate lattice QED in an external magnetic field at 
$\alpha=e^2/4\pi=1/137$, close to its physical value, over a range of $eB$
values with $m=0.1$ and $m=0.2$ on a $36^4$ lattice. To try and measure the
value of the condensate $\langle\bar{\psi}\psi\rangle$ in the $m=0$ limit,
we also simulate at a larger $\alpha$ value $\alpha=1/5$ and relatively
large $eB \approx 0.4848$ over a much wider range of masses. At this $\alpha$
we also perform simulations at $eB=0$ for comparison. At the smallest masses,
we plan larger lattice simulations needed to test for a non-zero condensate
at $m=0$. Very short exploratory runs on larger lattices suggest that the 
condensate does have a measurably large $m=0$ limit. In all our simulations we
store a configuration every 100 trajectories for further analysis.

\section{Lattice QED in an external magnetic field}

We simulate using the non-compact gauge action
$$
S(A) = \frac{\beta}{2}\sum_{n,\mu < \nu}[A_\nu(n+\hat{\mu})-A_\nu(n)
                                        -A_\mu(n+\hat{\nu})+A_\mu(n)]^2
$$
where $n$ is summed over the lattice sites, and $\mu$ and $\nu$ run from $1$ to
$4$ subject to the restriction. $\beta=1/e^2$. The expectation value of an
observable ${\cal O}(A)$ is then
$$
\langle{\cal O}\rangle = \frac{1}{Z}\int_{-\infty}^\infty \Pi_{n,\mu}
           dA_\mu(n) e^{-S(A)}[\det{\cal M}(A+A_{ext})]^{1/8}{\cal O}(A)
$$
where ${\cal M} = M^\dag M$, $A$ is the dynamic photon field and $A_{ext}$ is
the external photon field while
$$
M(A+A_{ext}) = \sum_\mu D_\mu(A+A_{ext})+m
$$
where the operator $D_\mu$ is defined by
\begin{eqnarray*}
[D_\mu(A+A_{ext})\psi](n) &=&
\frac{1}{2}\eta_\mu(n)\{e^{i(A_\mu(n)+A_{ext,\mu}(n))}\psi(n+\hat{\mu})\\
&-&e^{-i(A_\mu(n-\hat{\mu})+A_{ext,\mu}(n-\hat{\mu}))}\psi(n-\hat{\mu})\}
\end{eqnarray*}
and $\eta_\mu$ are the staggered phases.

We use the RHMC simulation method of Clark and Kennedy using rational
approximations to  ${\cal M}^{-1/8}$ and ${\cal M}^{\pm 1/16}$. To account for
the range of normal modes of the non-compact gauge action, we randomly vary
the trajectory lengths over the range of periods of these modes
\cite{Hands:1992uv}. $A_{ext}$ are chosen in the symmetric gauge in the x-y
plane so that the magnetic fields from each plaquette are in the +z-direction
and have the value $B$ modulo $2\pi$. This requires $eB=2\pi n/(n_1 n_2)$, where $n_1$ and $n_2$ are the lattice dimensions in the x and y directions, and
$n$ is an integer in the range $[0,n_1 n_2/2]$ \cite{Alexandre:2001pa}.

One of the observables we calculate is the electron contribution to the
effective gauge action per site $\frac{-1}{8V}{\rm trace}[\ln({\cal M})]$.
For this we use a rational approximation to $\ln$ following Kelisky and Rivlin
\cite{KR}, and a stochastic approximation to the trace. 

\section{Simulations and Results}

We first compare the chiral condensate per site for free electrons in an
external magnetic field,
$\langle\bar{\psi}\psi\rangle = {\rm trace}[M^{-1}(A_{ext})]/4V$, measured on a
$36^4$ lattice as a function of allowed $eB$ values and compare it with the
known continuum result:
$$
\langle\bar{\psi}\psi\rangle=\langle\bar{\psi}\psi\rangle|_{eB=0}
+\frac{meB}{4\pi^2}\int_0^\infty \frac{ds}{s}e^{-sm^2}\left[\coth(eBs)
  -\frac{1}{eBs}\right].
$$
for the `safe' mass values $m=0.1$ and $m=0.2$ \footnote{A safe mass value is
one for which $m << 1$ to keep discretization errors low, while $mL >> 1$
(where L is the linear size of the lattice) to keep finite lattice size effects
small.}.

We next calculate the effective fermion action per site,
${\cal L}_f = -\frac{1}{4V}\ln\{\det[M(A_{ext})]\}
            = -\frac{1}{4V}{\rm trace}\{\ln[M(A_{ext})]\}$, on the lattice
with the same parameters, and compare it with the known continuum values
$$
{\cal L}_f = {\cal L}_f|_{B=0}
+\frac{(eB)^2}{24\pi^2}\int_0^\infty\frac{ds}{s}e^{-m^2s} 
+\frac{eB}{8\pi^2}\int_0^\infty\frac{ds}{s^2}e^{-m^2s}\left[\coth(eBs)
  -\frac{1}{eBs}-\frac{eBs}{3}\right].
$$

\begin{figure}[htb]
\parbox{2.9in}{      
\epsfxsize=2.9in 
\epsffile{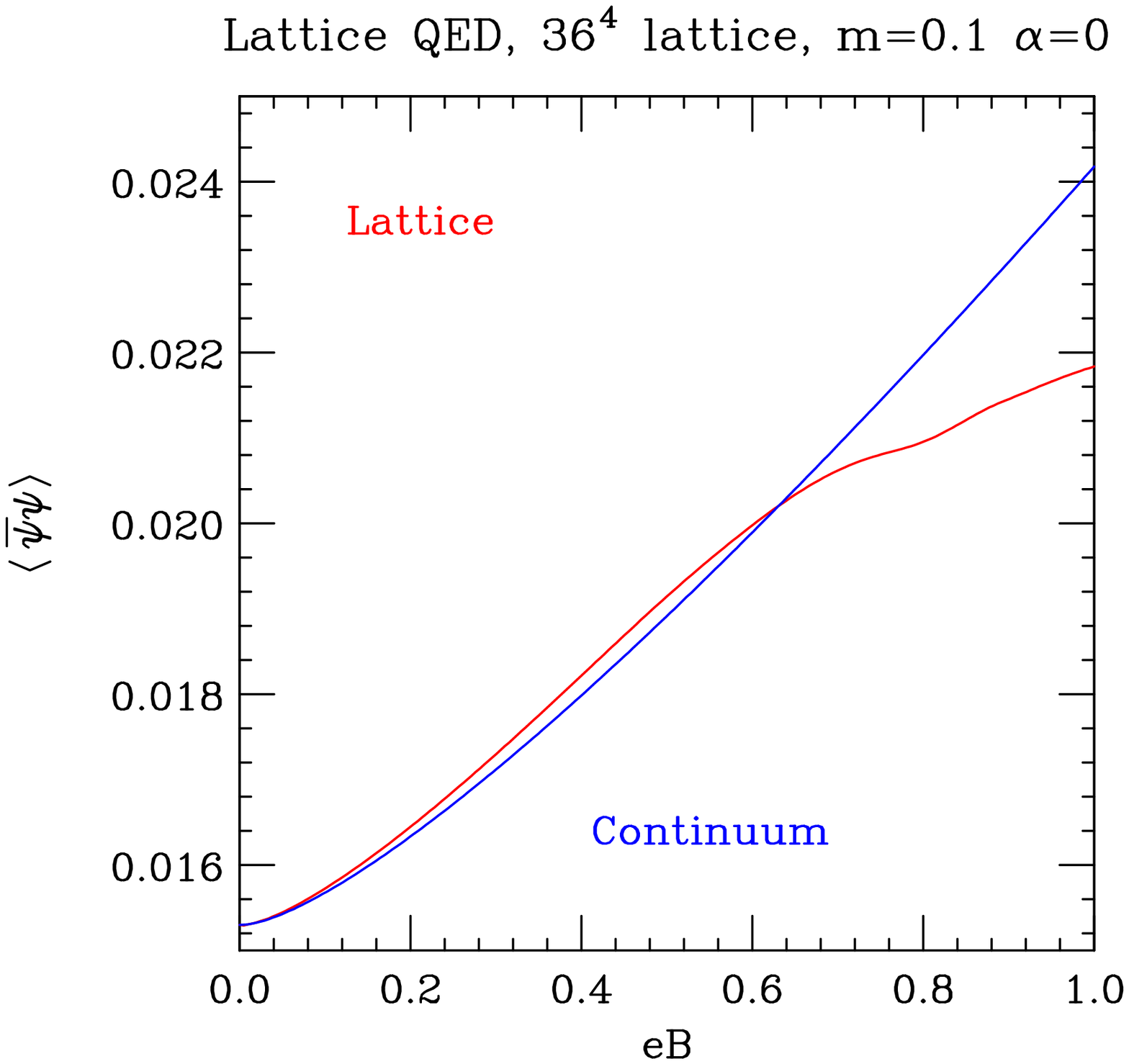}
\caption{Free-electron $\langle\bar{\psi}\psi\rangle$ as functions of $eB$,
comparing the continuum and lattice results for $m=0.1$.}
\label{fig:pbp_free}
}
\parbox{0.2in}{}
\parbox{2.9in}{                   
\epsfxsize=2.9in
\epsffile{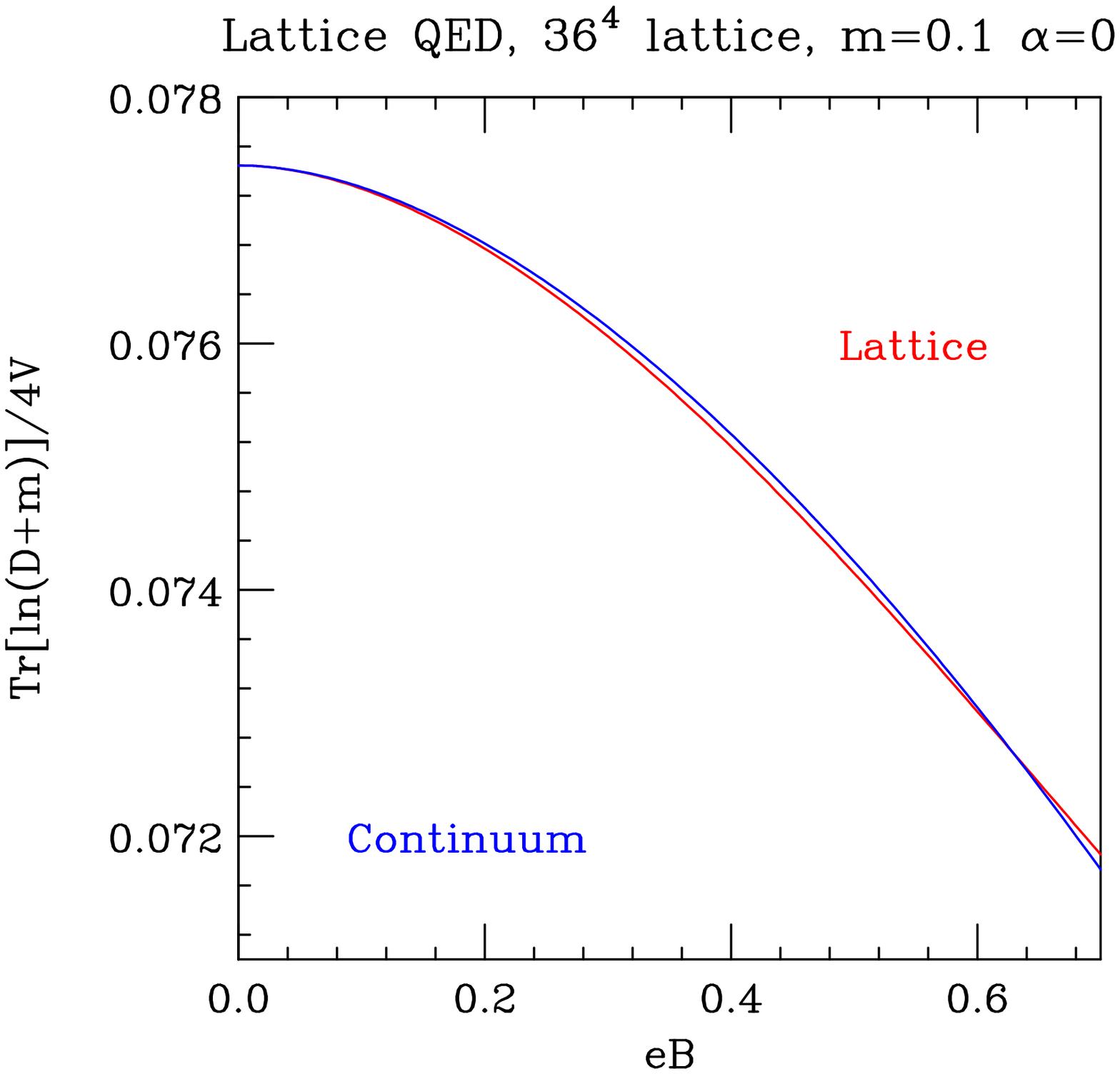}
\caption{Minus the effective action/site as functions of $eB$ for a free 
electron, comparing the continuum and lattice results for $m=0.1$.}
\label{fig:L_free}
}
\end{figure}

Figures~\ref{fig:pbp_free}, \ref{fig:L_free} show the chiral condensate and
effective fermion action as functions of $eB$, comparing the lattice results
on a $36^4$ lattice with the known continuum values. This indicates that we
have acceptable agreement for $eB \lesssim 0.65$.

\begin{figure}[htb]
\parbox{2.9in}{
\epsfxsize=2.9in
\centerline{\epsffile{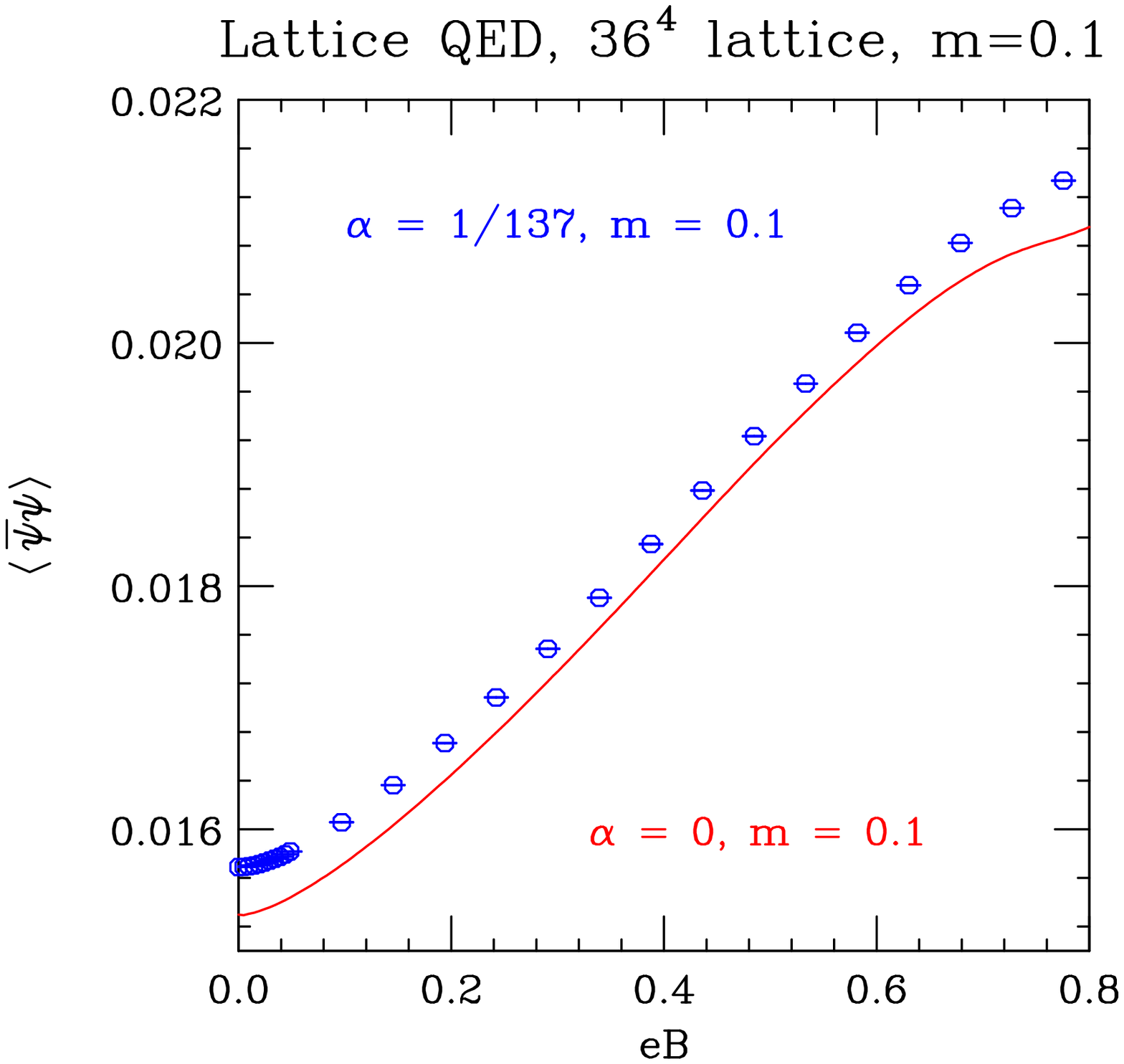}}
\caption{Electron $\langle\bar{\psi}\psi\rangle$ as functions of $eB$,
comparing the lattice free and QED results for $m=0.1$, $\alpha=1/137$} 
\label{fig:pbp}
}
\parbox{0.2in}{}
\parbox{2.9in}{                   
\epsfxsize=2.9in
\centerline{\epsffile{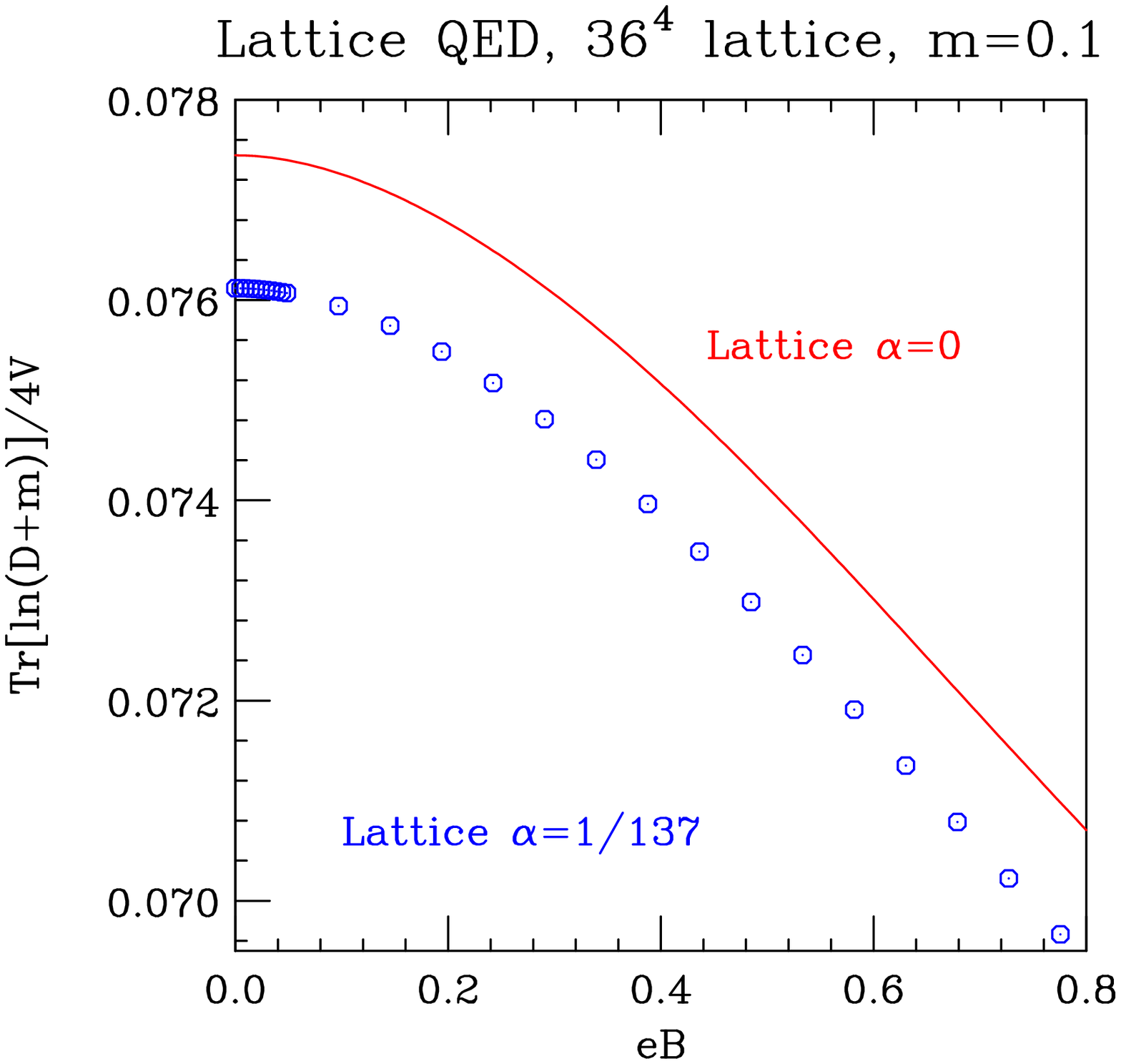}}
\caption{Minus the effective action/site as functions of $eB$, comparing the
lattice free and QED results for $m=0.1$, $\alpha=1/137$.}
\label{fig:logD}
}
\end{figure}

We simulate QED on a $36^4$ lattice with electron mass $m=0.1$ and $m=0.2$ in
an external magnetic field $B$, with $\alpha=1/137$, close to its physical
value. Note that here the momentum cutoff is so low that the difference between
bare and renormalized coupling and mass are at most a few percent and are
neglected. We simulate for 12,500 trajectories for each $m$ and $B$, storing
configurations every 100 trajectories for future analysis. Figure~\ref{fig:pbp}
shows the condensates $\langle\bar{\psi}\psi\rangle$ as functions of $eB$ for
these runs compared with those for free electrons for $m=0.1$. The most
noticeable effect is that QED increases the values of these condensates.
Figure~\ref{fig:logD} compares the effective fermion actions for QED with those
for free electrons.
Although the condensate decreases with decreasing mass, and is clearly small in
the limit $m \rightarrow 0$, these masses are not small enough to tell whether
it vanishes in this limit over the whole range of $eB$ as it does in the free
electron case, or approaches a finite value (except at $eB=0$) as approximate
calculations suggest.

To determine whether magnetic catalysis does occur requires simulating at
much smaller masses, where finite size effects are a potential problem.
However, if the condensate does vanish in the $m=0$ limit, the condensate is
presumably a large momentum effect as it is for free electrons in an external
magnetic field, and should be insensitive to finite size effects. If the
condensate remains finite as $m \rightarrow 0$, this is due to the presence of
near-zero modes. In this case, the condensate should show little finite lattice
size dependence for larger masses, and increase with increasing lattice size
for small masses. For large values of $eB$ where the system is expected to be
predominantly in the low lying Landau levels, the projection of the electron
orbits on the plane orthogonal to the magnetic field is small (dimensional
reduction). This suggests that to perform finite size analyses, one only needs
to extend the lattice in the direction of the magnetic field and the time
direction, while leaving the extent in the directions orthogonal to the
external field fixed. This reduces the resources needed to perform a finite
size analysis.

Since the zero mass condensate is still expected to be small, we need to do
everything in our power to increase its value. This could be done by increasing
the size of $eB$, however, the lattice restricts how far this can be increased.
We are already planning on reducing $m$. This leaves increasing $\alpha$ as
a method for increasing $\langle\bar{\psi}\psi\rangle$. For this reason, we
are now simulating at $\alpha=1/5$ to enhance the signal, and using electron
masses as low as $m=0.001$, still on a $36^4$ lattice. We are simulating at
$eB=2\pi\times 100/36^2\approx 0.4848$, reasonably large, but comfortably
below $eB\approx 0.65$, where lattice and continuum results are expected to
diverge. In addition, we are simulating at $\alpha=1/5$ and $eB=0$ for
comparison.

\begin{figure}[htb]
\parbox{2.9in}{
\epsfxsize=2.9in
\centerline{\epsffile{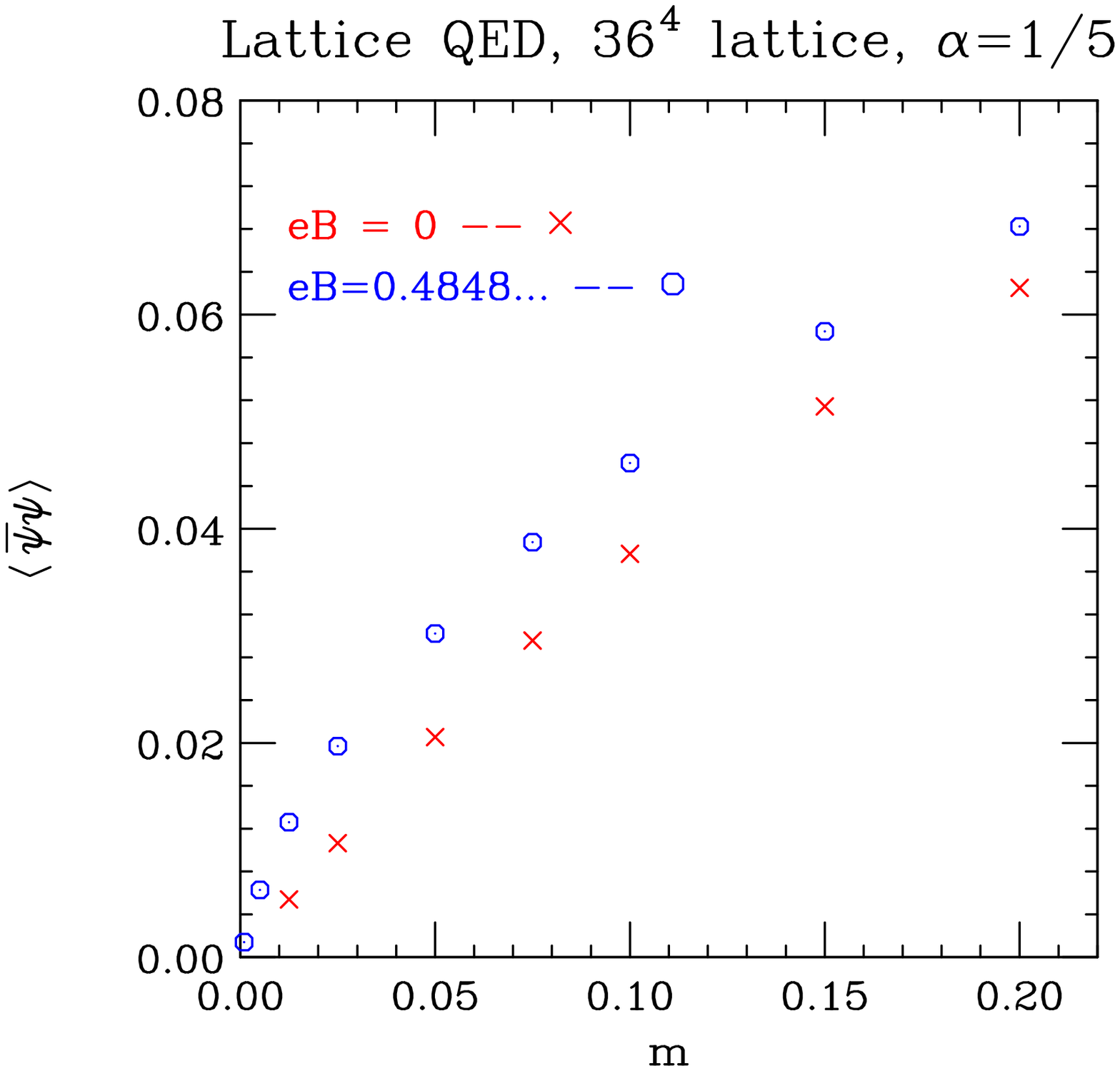}}
\caption{$\langle\bar{\psi}\psi\rangle$ as functions of $m$ for $eB=0$,
lower set of points, and $eB=2\pi\times 100/36^2$, upper set of points, for
$\alpha=1/5$ on a $36^4$ lattice.}
\label{fig:pbp_1/5}
}
\parbox{0.2in}{}
\parbox{2.9in}{
\epsfxsize=2.9in
\centerline{\epsffile{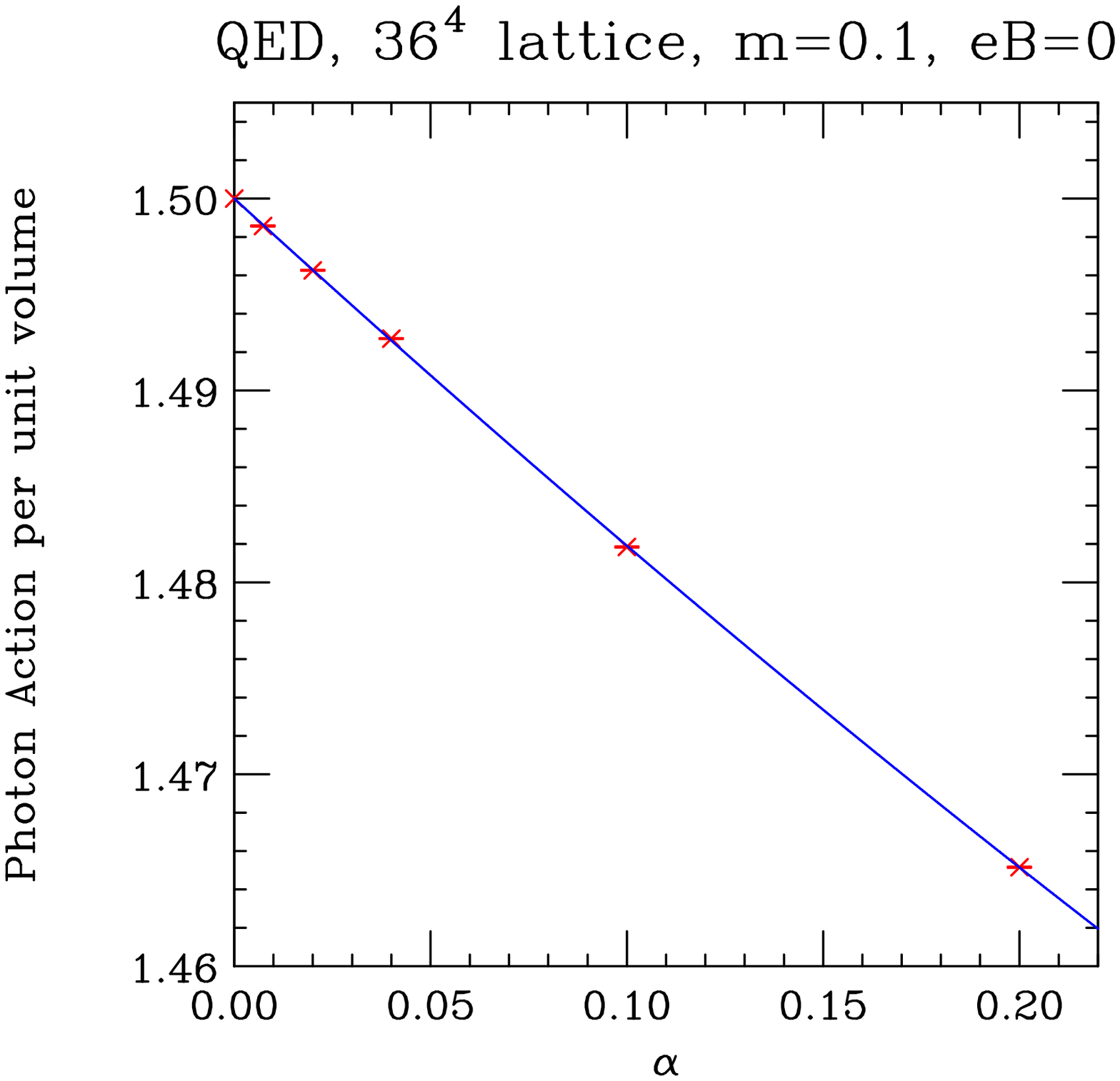}}
\caption{Gauge action/site at $eB=0$, $m=0.1$ on a $36^4$ lattice as a function
of $\alpha$. The blue curve is a fit quadratic in $\alpha$.} 
\label{fig:plaquette}
}
\end{figure}

Figure~\ref{fig:pbp_1/5} shows the condensates $\langle\bar{\psi}\psi\rangle$
as functions of mass $m$ from our simulations at $\alpha=1/5$ at $eB=0$ and at
$eB=2\pi\times 100/36^2$ on a $36^4$ lattice. While both appear to be
approaching zero as $m \rightarrow 0$, the $eB=0$ points do so relatively
smoothly over the whole range of $m$ values, while the points for non-zero
$eB$ and $m \ge 0.025$ appear to be headed for a non-zero value at $m=0$.
Below $m=0.025$ the points at non-zero $eB$, curve more strongly, approaching
something much closer to zero. This suggests that we need to perform a finite
size analysis at these lower mass values. Very preliminary, short, exploratory
simulations on $36^2 \times 72^2$ lattices at $\alpha=1/5$ and
$eB=2\pi\times 100/36^2$ appear to indicate that, while the condensate for
$m=0.025$ shows little if any change from its value on a $36^4$ lattice, those
for smaller masses show significant increases. In fact, for the smallest mass
($m=0.001$), the condensate on a $36^2 \times 64^2$ lattice appears to be
roughly twice that on a $36^4$ lattice, while that on a $36^2 \times 96^2$
lattice appears to be roughly thrice that on a $36^4$ lattice. If these
observations are ratified by simulations of significant length, we are seeing
the first evidence for magnetic catalysis for QED in an external magnetic field
from lattice QED simulations.

In figure~\ref{fig:plaquette} we plot the gauge(photon) action/site at $eB=0$,
$m=0.1$ on a $36^4$ lattice as a function of $\alpha$. The curve is the
quadratic fit $1.5-0.187499\,\alpha+0.0661374\,\alpha^2$, which has
$\chi^2/DOF$ of $11$. Since the ${\cal O}(\alpha)$ term dominates the QED
contribution, it suggests that one can probably trust perturbation theory for
this quantity. It should be checked against a perturbative calculation with
lattice regularization, which would provide a check of lattice simulations.

\section{Summary, Discussion and Conclusions}

We simulate lattice QED in a constant external magnetic field $B$ using the
RHMC method. For electrons in such an external field without QED, we measure
the chiral condensate and effective action over the allowed range of $eB$
($|eB| \le \pi$) and determine that the range of $eB$ over which the lattice
measurements agree with the known continuum values is $|eB| \lesssim 0.65$.

We perform simulations at the near-physical value of $\alpha$, namely at
$\alpha=1/137$, with electron masses $m=0.1$ and $m=0.2$ on a $36^4$ lattice,
over the range of $eB$ where lattice measurements are expected to be close to
their continuum values. The condensate $\langle\bar{\psi}\psi\rangle$
is consistently larger than that without QED and increases with increasing
$eB$. The effective action, which measures the response of the system to the
applied magnetic field also lies above that without QED and has a similar $eB$
dependence to that for the free electron in a magnetic field.

We also simulate at stronger coupling -- $\alpha=1/5$ -- over a wider range
of masses $0.001 \le m \le 0.2$ at $eB \sim 0.5$ and $eB=0$, also on a $36^4$
lattice, to determine whether having a (large) external magnetic field
combined with the electron-positron attraction due to QED is sufficient to
produce a non-zero condensate in the $m \rightarrow 0$ limit (magnetic
catalysis). Since the smaller mass values are far below what is considered
safe, we need to perform a finite size analysis to determine whether this
occurs. Because strong magnetic fields restrict the extent of the electron
wave-functions transverse to the external magnetic field we only need to
increase the lattice extent in the direction of the magnetic field and that of
time for this analysis. Very limited exploratory runs on such larger lattices
appear to indicate that $\langle\bar{\psi}\psi\rangle$ does indeed remain
finite in the zero mass limit.

We need to extend our simulations on larger lattices to test if
$\langle\bar{\psi}\psi\rangle$ is non-vanishing in the zero mass limit for
large $eB$ (and $\alpha$). If so, we should try to determine the $eB$ and
$\alpha$ dependence of this condensate.

We will use the configurations stored during these runs to determine the
effect of the external magnetic field and QED on the coulomb field of a
static point charge. In addition, we will measure the electron propagator on
these configurations to measure the dynamic contribution to the electron
mass.

Our next project will be to use lattice methods to study QED in an external
electric field. This is a much more difficult task, since the external electric
field makes the euclidean action complex. [It is the imaginary part of this
action which describes the instability of the vacuum to decay, producing
electron-positron pairs.] This means that standard simulation methods based on
importance sampling will fail. Those methods which remain are less reliable.

\acknowledgments

DKS's research is supported in part U.S. Department of Energy, Division of
High Energy Physics, under Contract No. DE-AC02-06CH11357. The high performance
computing was provided by the LCRC at Argonne National Laboratory on their
Bebop cluster. Access to Stampede-2 at TACC, Expanse at UCSD and Bridges-2 at
PSC was provided under an XSEDE allocation. Time on Cori at NERSC was provided
through an ERCAP allocation. DKS thanks G.~T.~Bodwin for inciteful discussions,
while JBK would like to thank V.~Yakimenko for discussions which helped inspire
this project, and I.~A.~Shovkovy for helpful discourse on magnetic catalysis.

\end{document}